\documentclass[journal]{IEEEtran}
\usepackage{etex}
\usepackage{stfloats}
\usepackage{multirow}

\usepackage{pstricks}
\usepackage{lipsum}
\usepackage[numbers,sort&compress]{natbib}
   
\usepackage{amssymb}
\usepackage{pgfplots}
\usepackage{graphicx}
\pgfplotsset{compat=1.3}
\usepackage[cmex10]{amsmath}
\usepackage{color}

\begin{document}
\title{The $\kappa$-$\mu$ Shadowed Fading Model: \\ Unifying the $\kappa$-$\mu$ and $\eta$-$\mu$ Distributions}
\hyphenation{hyper-geo-met-ric shad-ow-ing}

\author{\IEEEauthorblockN{Laureano Moreno-Pozas, F. Javier Lopez-Martinez, Jos\'e F. Paris and  Eduardo Martos-Naya}
\thanks{This work has been funded by the Consejer\'ia de Econom\'ia,
Innovaci\'on, Ciencia y Empleo of the Junta de Andaluc\'ia, the Spanish Government and the European Regional Development Fund (projects P2011-TIC-7109, P2011-TIC-8238, TEC2011-25473 and COFUND2013-40259), and also by the University of M\'alaga and European Union under Marie-Curie COFUND U-mobility program (ref.246550). The authors are with Departmento de Ingenier\'ia de Comunicaciones, Universidad de M\'alaga - Campus de Excelencia Internacional Andaluc\'ia
Tech., M\'alaga 29071, Spain (e-mail: lmp@ic.uma.es).}}

\maketitle

\begin{abstract}
This paper shows that the recently proposed $\kappa$-$\mu$ shadowed fading model includes, besides the $\kappa$-$\mu$ model, the {$\eta$-$\mu$} fading model as a particular case. This has important {relevance} in practice, as it allows for the unification of these popular fading distributions through a more general, yet equally tractable, model. The convenience of new underlying physical models is discussed. Then, we derive simple and novel closed-form expressions for the asymptotic ergodic capacity in $\kappa$-$\mu$ shadowed fading channels, which illustrate the effects of the different fading parameters on the system performance. By exploiting the unification here unveiled, the asymptotic capacity expressions for the $\kappa$-$\mu$ and $\eta$-$\mu$ fading models are also obtained in closed-form as special cases.
\end{abstract}

\section{Introduction}

The scientific community has been recently interested in the definition of new generalized fading models, aiming to provide a better fit to real measurements observed in different scenarios \cite{Durgin, Shankar, Yacoub, Yacoubalphamu, Paris}. In such context, the $\kappa$-$\mu$ and $\eta$-$\mu$ fading models \cite{Yacoub} have become very popular in the literature due to their versatility to accommodate to different propagation conditions and their relatively simple tractable form \cite{wang, peppas_eta, Yacoubalgorithm, peppas, sofotasios}.

The $\kappa$-$\mu$ and $\eta$-$\mu$ fading models, first introduced in \cite{Yacoub_kappa} and \cite{Yacoub_eta}, were independently derived to characterize very different propagation conditions. On the one hand, the $\kappa$-$\mu$ distribution can be regarded as a generalization of the classic Rician fading model for line-of-sight (LOS) scenarios, extensively used in spatially homogeneous propagation environments. On the other hand, the $\eta$-$\mu$ distribution can be considered as a generalization of the classic Nakagami-$q$ (Hoyt) fading model for non-LOS scenarios, often used in non-homogeneous environments. Therefore, and because they arise from different underlying physical models, there is no clear connection between the $\kappa$-$\mu$ and $\eta$-$\mu$ fading models.

One of the most appealing properties of the $\kappa$-$\mu$ and \mbox{$\eta$-$\mu$} fading models is that they include most popular fading distributions as particular cases. For instance, the Rician, Nakagami-$m$, Rayleigh and one-sided Gaussian models can be derived from the $\kappa$-$\mu$ fading by setting the parameters $\kappa$ and $\mu$ to specific real positive values. Similarly, the $\eta$-$\mu$ fading model includes the Nakagami-$q$, Nakagami-$m$, Rayleigh and one-sided Gaussian as special cases. 

Very recently, the $\kappa$-$\mu$ shadowed fading model was introduced in \cite{Paris}, with the aim of jointly including large-scale and small-scale propagation effects. This new model exhibits excellent agreement when compared to measured land-mobile satellite \cite{Alouini_land_mobile}, underwater acoustic \cite{Paris, ParisConf} and body communications fading channels \cite{Cotton,Cotton2}, by considering that the dominant components are affected by random fluctuations \cite{Paris}. This model \cite{Paris} includes the popular Rician shadowed fading distribution \cite{Alouini_land_mobile} as a particular case, and obviously it also includes the $\kappa$-$\mu$ fading distribution from which it originates. However, as we will later see, the versatility of the $\kappa$-$\mu$ shadowed fading model has not been exploited to the full extent possible.

In this paper we show that the $\kappa$-$\mu$ shadowed distribution unifies the set of homogeneous fading models associated with the $\kappa$-$\mu$ distribution, and strikingly, it also unifies the set of non-homogeneous fading models associated with the $\eta$-$\mu$ distribution, { which may seem counterintuitive at first glance}. In addition to a formal mathematical proof of how the {main} probability functions introduced by Yacoub originate from the ones derived in \cite{Paris}, we also establish new underlying physical models for the $\kappa$-$\mu$ shadowed distribution that justify these phenomena. In fact, we propose a novel method to derive the Nakagami-$q$ (Hoyt) and the $\eta$-$\mu$ distributions which consists in using the shadowing of the dominant components to recreate a non-homogeneous propagation environment. This connection, which is here proposed for the first time in the literature, has important implications in practice: first, and contrary to the common belief, it shows that the $\kappa$-$\mu$ and $\eta$-$\mu$ fading distributions are connected. Hence we can jointly study the \mbox{$\kappa$-$\mu$} and $\eta$-$\mu$ fading models by using a common approach instead of separately \cite{Yacoub, Ermolova, Annamalai}. Besides, it implies that when deriving any performance metric for the $\kappa$-$\mu$ shadowed fading model, we are actually solving the same problem for the simpler $\kappa$-$\mu$ and $\eta$-$\mu$ distributions at no extra cost. 

Leveraging our novel approach, we derive simple and closed-form asymptotic expressions for the ergodic capacity of communication systems operating under $\kappa$-$\mu$ shadowed fading in the high signal-to-noise ratio regime, which can be evidently employed  for the $\kappa$-$\mu$ and $\eta$-$\mu$ distributions. Unlike the exact analyses in \cite{daCosta} and \cite{Celia} which require the use of the Meijer G- and bivariate Meijer G-functions, our results allow for a better insight into the effects of the fading parameters on the capacity.

The remainder of this paper is structured as follows. In section II, we introduce the notation, as well as some definitions and preliminary results. In section III, we propose new physical models for the $\kappa$-$\mu$ shadowed distribution. In section IV we show how the $\kappa$-$\mu$ and $\eta$-$\mu$ distributions naturally arise as particular cases of the $\kappa$-$\mu$ shadowed fading model. In section V, we use these results to investigate the ergodic capacity in \mbox{$\kappa$-$\mu$} shadowed fading channels and thus for the $\kappa$-$\mu$ and \mbox{$\eta$-$\mu$} channels. In section VI, numerical results are presented. Finally conclusions are drawn.

\section{Notation and Preliminaries}

Throughout this paper, we differentiate the complex from the real random variables by adding them a tilde on top, so that $x$ is a real random variable and $\widetilde{z}$ is a complex random variable. $\mathbb{E}[\cdot]$ is the expectation operator. The symbol $\sim$ signifies \emph{statistically distributed as} while i.i.d means \emph{independent and identically distributed}. {Thus, $x\sim\mathcal{N}(\bar{x},\sigma^2)$ symbolizes that $x$ is statically distributed as a real Gaussian random variable with mean $\bar{x}$ and variance $\sigma^2$, while $\widetilde{z}\sim\mathcal{CN}(\bar{z},\sigma^2)$ means that $\widetilde{z}$ is statistically distributed as a circularly symmetric complex Gaussian random variable with mean $\bar{z}$ and variance $\sigma^2$.} 
Moreover, although the classic fading distributions are usually defined in the literature by its envelope probability density function (pdf), we here present the results in terms of power probability density functions.

Next, we present some definitions which are employed in the paper.

\emph{Definition 1: Higher-order amount of fading} \cite{Yilmaz}.\\
The $n$th-order amount of fading $AF_{\gamma}^{(n)}$ is defined as
\begin{equation}
\label{AF}
AF_{\gamma}^{(n)}=\frac{\mathbb{E}[\gamma^n]}{\bar{\gamma}^n}-1
\end{equation}
for the instantaneous SNR $\gamma$, where $\bar{\gamma}^n=\mathbb{E}[\gamma]^n$.

\emph{Definition 2: The generalized hypergeometric function}.\\
The generalized hypergeometric function of one scalar argument is defined as
\begin{equation}
\label{hyper}
_p\mathcal{F}_q(a_1,\ldots,a_p;b_1,\ldots,b_q;x)=\sum_{r=0}^{+\infty}\frac{(a_1)_r\ldots(a_p)_r}{(b_1)_r\ldots(b_q)_r}\frac{x^r}{r!},
\end{equation}
where $(a)_r$ is the Pochhammer symbol \cite[eq. (6.1.22)]{Abramowitz}, ${a_i\in\mathbb{C}}$ and $b_j\in\mathbb{C}^*\backslash\mathbb{Z}^-$.

\emph{Definition 3: The gamma distribution}.\\
Let $w$ be a random variable which statistically follows a gamma distribution with shape parameter $\alpha$ and rate parameter $\beta$, i.e, $w\sim\Gamma(\alpha,\beta)$, then its pdf is given by
\begin{equation}
\label{pdf_Gamma}
f_w(w)=\frac{\beta^{\alpha}}{\Gamma(\alpha)}w^{\alpha-1}e^{-\beta w},
\end{equation}
where $\Gamma(\cdot)$ is the gamma function \cite[eq. (6.1.1)]{Abramowitz} and ${\alpha,\beta\in\mathbb{R}^+}$.

\emph{Definition 4: The $\kappa$-$\mu$ shadowed distribution} \cite{Paris}.\\
Let $\gamma$ be a random variable which statistically follows a \mbox{$\kappa$-$\mu$} shadowed distribution with mean $\bar{\gamma}=\mathbb{E}[\gamma]$ and non-negative real shape parameters $\kappa$, $\mu$ and $m$, i.e, ${\gamma\sim\mathcal{S}_{\kappa\mu m}(\bar{\gamma}; \kappa, \mu, m)}$, then its pdf is given by
\begin{equation}
\label{pdf}
\begin{split}
f_{\gamma}(\gamma)=&\frac{\mu^\mu m^m(1+\kappa)^\mu}{\Gamma(\mu)\bar{\gamma}(\mu\kappa+m)^m}\left(\frac{\gamma}{\bar{\gamma}}\right)^{\mu-1}\\
&\times e^{-\frac{\mu(1+\kappa)\gamma}{\bar{\gamma}}}\ _1\mathcal{F}_1\left(m;\mu;\frac{\mu^2\kappa(1+\kappa)}{\mu\kappa+m}\frac{\gamma}{\bar{\gamma}}\right),
\end{split}
\end{equation}
where $_1\mathcal{F}_1(\cdot)$ is the confluent hypergeometric function of scalar argument \cite[eq. (13.1.2)]{Abramowitz}, which is a particular case of eq.~(\ref{hyper}).

\emph{Definition 5: The $\kappa$-$\mu$ distribution} \cite{Yacoub}.\\
Let $\gamma$ be random variable which statistically follows a $\kappa$-$\mu$ distribution with mean $\bar{\gamma}=\mathbb{E}[\gamma]$ and non-negative real shape parameters $\kappa$ and $\mu$, i.e, $\gamma\sim\mathcal{S}_{\kappa\mu}(\bar{\gamma}; \kappa, \mu)$, then its pdf is given by
\begin{equation}
\label{pdf_kappa_mu}
\begin{split}
f_{\gamma}(\gamma)=&\frac{\mu(1+\kappa)^\frac{\mu+1}{2}}{\bar{\gamma}\kappa^\frac{\mu-1}{2}e^{\mu\kappa}}\left(\frac{\gamma}{\bar{\gamma}}\right)^{\frac{\mu-1}{2}}\\
&\times e^{-\frac{\mu(1+\kappa)\gamma}{\bar{\gamma}}}I_{\mu-1}\left(2\mu\sqrt{\frac{\kappa(1+\kappa)\gamma}{\bar{\gamma}}}\right),
\end{split}
\end{equation}
where $I_k(\cdot)$ is the $k$-th order modified Bessel function of first kind, which can be defined in terms of the Bessel hypergeometric function $_0\mathcal{F}_1(\cdot)$ \cite[eq. (9.6.47)]{Abramowitz}.

\emph{Definition 6: The $\eta$-$\mu$ distribution (Format 1)} \cite{Yacoub}.\\
Let $\gamma$ be random variable which statistically follows an $\eta$-$\mu$ distribution with mean $\bar{\gamma}=\mathbb{E}[\gamma]$ and non-negative real shape parameters $\eta$ and $\mu$, i.e, $\gamma\sim\mathcal{S}_{\eta\mu}(\bar{\gamma}; \eta, \mu)$, then its pdf is given by
\begin{equation}
\label{pdf_eta_mu}
\begin{split}
f_{\gamma}(\gamma)=&\frac{\sqrt{\pi}(1+\eta)^{\mu+\frac{1}{2}}\mu^{\mu+\frac{1}{2}}}{\Gamma(\mu)\bar{\gamma}\sqrt{\eta}(1-\eta)^{\mu-\frac{1}{2}}}\left(\frac{\gamma}{\bar{\gamma}}\right)^{\mu-\frac{1}{2}}\\
&\times e^{-\frac{\mu(1+\eta)^2\gamma}{2\eta\bar{\gamma}}}I_{\mu-\frac{1}{2}}\left(\frac{\mu(1-\eta)^2}{2\eta}\frac{\gamma}{\bar{\gamma}}\right).
\end{split}
\end{equation}

\section{Appropriate Physical Models}

In this section, we discuss the underlying physical model of the $\kappa$-$\mu$ shadowed presented in \cite{Paris}. Then we propose two more general and suitable physical models which will allow us to later unify the $\kappa$-$\mu$ and $\eta$-$\mu$ distributions.

\subsection{Previous model}
The $\kappa$-$\mu$ shadowed fading is presented as the generalization of the $\kappa$-$\mu$ model in \cite{Paris}. Like in the $\kappa$-$\mu$ fading case, the signal can be decomposed into different clusters of waves, where each cluster has scattered waves with similar delays and the delay spreads of different clusters are relatively large \cite{Yacoub}. In the \mbox{$\kappa$-$\mu$} shadowed fading, the difference appears in the dominant component of each cluster, which is no longer deterministic and can randomly fluctuate because of shadowing. In fact, when we consider that the dominant component of each cluster suffers from the same shadowing, we can express the total signal power in terms of the in-phase and the quadrature components of each cluster as \cite{Paris}
\begin{equation}
\label{model_Paris}
W=\sum_{i=1}^{\mu}(x_i+\xi p_i)^2+(y_i+\xi q_i)^2,
\end{equation} 
where $\mu$ is the number of clusters; $x_i$ and $y_i$ are independent real Gaussian random variables with zero mean and variance $\sigma^2$, which represent the real and imaginary parts of the \mbox{$i$-th} cluster scattering waves; $p_i^2$ and $q_i^2$ are the powers of the real and imaginary parts of the $i$-th dominant component, respectively, and $\xi$ is the shadowing component which is statistically distributed as a Nakagami-$m$ random variable with shape parameter $m$ and $\mathbb{E}[\xi^2]=1$.

It is known that this physical model follows a $\kappa$-$\mu$ shadowed distribution \cite{Paris}, i.e., the instantaneous signal-to-noise ratio (SNR) $\gamma=\bar{\gamma}W/\bar{W}$, with $\bar{\gamma}=\mathbb{E}[\gamma]$, $\bar{W}=\mathbb{E}[W]=d^2+2\sigma^2\mu$ and $d^2=\sum_{i=1}^{\mu}p_i^2+q_i^2$, is distributed as $\mathcal{S}_{\kappa\mu m}(\bar{\gamma}; \kappa, \mu, m)$, where the parameter ${\kappa=d^2/{2\sigma^2\mu}}$ represents the ratio between the total power of the dominant components and the total power of the scattered waves. It is worth noticing that this result can be extended for $\mu$ taking {non-integer} positive values \cite{Paris}, despite the model loses its physical meaning. 

However, this model imposes that the shadowing $\xi$ is statistically distributed as a Nakagami-$m$ random variable, which is a strict condition that is relaxed in the next section. 

\subsection{Generalized model with the same shadowing for all the clusters}

{The previous model in eq.~(\ref{model_Paris}) clearly separates each cluster in the real and imaginary power components, so that the model can be defined by only using real random variables. Thus, using this time complex random variables, it can be reformulated as
\begin{equation}
\label{model_paris_reformulated}
W=\sum_{i=1}^{\mu}\vert\widetilde{z}_i+\xi\widetilde{\rho}_i\vert^2,
\end{equation}
where $\widetilde{z}_i$ and $\widetilde{\rho}_i$ can be related to the variables of the previous model in form of $\widetilde{z}_i=x_i+jy_i$ and $\widetilde{\rho}_i=p_i+jq_i$. Hence, $\widetilde{z}_i\sim\mathcal{CN}(0,2\sigma^2)$ represents the scattering wave of the $i$-th cluster and $\vert\widetilde{\rho}_i\vert^2$ is the deterministic dominant power of the $i$-th cluster.

A straightforward generalization of the previous model is to consider a complex shadowing component, so that we obtain the following model
\begin{equation}
\label{modelo_general_Paris}
\Omega_1=\sum_{i=1}^{\mu}\vert\widetilde{z}_i+\widetilde{\xi}\widetilde{\rho}_i\vert^2,
\end{equation}
where $\widetilde{\xi}$ is now a complex random variable, with ${\vert\widetilde{\xi}\vert^2\sim\Gamma(m,m)}$ and arbitrary phase.

This new model obviously represents a similar scenario as the previous one in Section III.A, since all the clusters suffer from the same shadowing $\widetilde{\xi}$, which can be justified by the fact that the shadowing can occur near the transmitter or receiver side. The pdf of the instantaneous SNR of the model in eq.~(\ref{modelo_general_Paris}) is derived as follows.

\emph{Lemma 1}: Let $\gamma_1=\bar{\gamma}_1\Omega_1/\bar{\Omega}_1$, with $\bar{\gamma}_1=\mathbb{E}[\gamma_1]$, be the instantaneous SNR of the model in eq.~(\ref{modelo_general_Paris}). Then ${\gamma_1\sim\mathcal{S}_{\kappa\mu m}(\bar{\gamma}_1; \kappa, \mu, m)}$ where $\kappa=\sum_{i=1}^{\mu}\vert\widetilde{\rho}_i\vert^2/2\sigma^2\mu$.

\emph{Proof}: The signal power $\Omega_1$ conditioned to the shadowing power $\vert\widetilde{\xi}\vert^2$ follows a $\kappa$-$\mu$ distribution. Moreover, since ${\vert\widetilde{\xi}\vert^2\sim\Gamma(m,m)}$, then $\vert\widetilde{\xi}\vert^2\sum_{i=1}^{\mu}\vert\widetilde{\rho}_i\vert^2\sim\Gamma(m,m\sum_{i=1}^{\mu}\vert\widetilde{\rho}_i\vert^2)$. Thus, the conditional form here obtained is equivalent to the one in \cite{Paris} and so we can follow the same steps to prove that the SNR of the model follows a $\kappa$-$\mu$ shadowed distribution.

In fact, we have proved that the distribution of the model is independent of the phase of the shadowing component $\widetilde{\xi}$. In the next section we propose another physical model for the $\kappa$-$\mu$ shadowed distribution.}

\subsection{Generalized model with i.i.d. shadowing}

When the shadowing does not occur near the transmitter or receiver sides, all the clusters could suffer from different shadowing effects, so the instantaneous received power can be expressed as 
\begin{equation}
\label{model_propuesto}
\Omega_2=\sum_{i=1}^{\mu}\vert\widetilde{z}_i+\widetilde{\xi}_i \widetilde{\rho}\vert^2,
\end{equation} 
where { $\widetilde{z}_i\sim\mathcal{CN}(0,2\sigma^2)$}; $\vert\widetilde{\rho}\vert^2$ is the power of each dominant component; and $\widetilde{\xi}_i$ is a complex random variable which represents the shadowing component of the $i$-th cluster, where {$\widetilde{\xi}_i$ are i.i.d., i.e, ${\vert\widetilde{\xi}_i\vert^2\sim\Gamma(\hat{m},\hat{m})\ \forall i}$}. 

{Actually}, these propagation conditions are likely to occur in real scenarios where the dominant components of different clusters could travel through different paths that are separated enough to suffer from independent large-scale propagation effects. The pdf of the instantaneous SNR of the model in eq.~(\ref{model_propuesto}) is derived next.

\emph{Lemma 2}: Let $\gamma_2=\bar{\gamma}_2\Omega_2/\bar{\Omega}_2$, with $\bar{\gamma}_2=\mathbb{E}[\gamma_2]$, be the instantaneous SNR of the model in eq.~(\ref{model_propuesto}). Then ${\gamma_2\sim\mathcal{S}_{\kappa\mu m}(\bar{\gamma}_2; \kappa, \mu, m)}$ where $\kappa=\vert\widetilde{\rho}\vert^2/2\sigma^2\mu$ and {$m=\mu\cdot\hat{m}$}.

\emph{Proof}: The conditioned signal power $\Omega_2$ conditioned to the sum of the i.i.d. shadowed dominant component powers $P=\vert\widetilde{\rho}\vert^2\sum_{i=1}^{\mu}\vert\widetilde{\xi}_i\vert^2$ follows a $\kappa$-$\mu$ distribution. Moreover, since $\vert\widetilde{\xi}_i\vert^2\sim\Gamma(\hat{m},\hat{m})\ {\forall i}$, then $P\sim\Gamma(m,\vert\widetilde{\rho}\vert^2m)$, where $m=\sum_{i=1}^\mu \hat{m}$. Thus, again we have the same conditional form as in \cite{Paris} and so we can follow the same steps to prove that the SNR of the model follows a $\kappa$-$\mu$ shadowed distribution. {\footnote{{The different shadowing components $\widetilde{\xi}_i$ do not have to be identically distributed to complete the proof. All that is needed is that the normalized rate parameter $\beta_i/m_i$ of each shadowing power $\vert\widetilde{\xi}_i\vert^2$ must be equal $\forall i$. Although from a mathematical point of view this is a valid model, this scenario is hard to imagine in practical conditions. Therefore, we will restrict ourselves to the case with i.i.d. shadowing components.}}}

Therefore, the SNRs of both physical models presented in eq.~(\ref{modelo_general_Paris}) and eq.~(\ref{model_propuesto}) follow a $\kappa$-$\mu$ shadowed distribution. The closed-form expressions for the cumulative distribution and the moment generating functions can be found in \cite{Paris}.

\section{$\kappa$-$\mu$ and $\eta$-$\mu$ Unification}
In the previous section, we have introduced two different physical models which lead to the $\kappa$-$\mu$ shadowed distribution. Now, we show how each of these models reduces to the general $\kappa$-$\mu$ and $\eta$-$\mu$ fading distributions, respectively. By doing so, we show that the $\kappa$-$\mu$ shadowed distribution can unify all classic fading models, both for homogeneous and non-homogeneous propagation conditions, and their most general counterparts \cite{Yacoub}. 

\subsection{$\kappa$-$\mu$ distribution and particular cases}
The $\kappa$-$\mu$ distribution is destined to model homogeneous environments, where the scattering for each cluster can be modeled with a circularly symmetric random variable. The derivation of the $\kappa$-$\mu$ distribution from Lemma 1 is given in the following corollary.

\emph{Corollary 1}: Let $\gamma_1=\bar{\gamma}_1\Omega_1/\bar{\Omega}_1$, with $\bar{\gamma}_1=\mathbb{E}[\gamma_1]$, be the instantaneous SNR of the model in eq.~(\ref{modelo_general_Paris}), i.e., ${\gamma_1\sim\mathcal{S}_{\kappa\mu m}(\bar{\gamma}_1; \kappa, \mu, m)}$. If $m\rightarrow\infty$, $\gamma_1\sim\mathcal{S}_{\kappa\mu}(\bar{\gamma}_1; \kappa, \mu)$.

\emph{Proof}: By taking the limit $m\rightarrow\infty$ in eq.~(\ref{pdf}) and applying the following properties
\begin{equation}
\label{limite3}
\lim_{a\rightarrow\infty}~_1\mathcal{F}_1\Big(a; b; \frac{1}{a}z\Big)=~_0\mathcal{F}_1\Big(b; z\Big),
\end{equation}
\begin{equation}
\label{limite_m}
\lim_{a\rightarrow \infty}~\left(1+\frac{1}{a}x\right)^{-a}=e^{-x},
\end{equation}
where the eq.~(\ref{limite_m}) is the well-known limit that defines the exponential function, we obtain the pdf in eq.~(\ref{pdf_kappa_mu}).

Corollary 1 is interpreted as follows: the $\kappa$-$\mu$ distribution is derived by eliminating completely the shadowing of each dominant component, which can be done by taking $m\rightarrow\infty$, so that the dominant component of each cluster becomes deterministic. Actually, as the parameter $m$ grows, the pdf of each dominant component is gradually compressed and, at the limit $m\rightarrow\infty$, it becomes a Dirac delta function. Thus, the model is defined by a  circularly symmetric complex random variable with some non-zero mean in each cluster, so that we obtain the $\kappa$-$\mu$ model, whereas in case that $\mu=1$ we have the Rician fading model.

Next, we derive the Nakagami-$m$ physical model from the physical model in eq.~(\ref{modelo_general_Paris}).

\emph{Corollary 2}: Let $\gamma_1=\bar{\gamma}_1\Omega_1/\bar{\Omega}_1$, with $\bar{\gamma}_1=\mathbb{E}[\gamma_1]$, be the instantaneous SNR of the model in eq.~(\ref{modelo_general_Paris}), i.e., ${\gamma_1\sim\mathcal{S}_{\kappa\mu m}(\bar{\gamma}_1; \kappa, \mu, m)}$. If $\kappa\rightarrow 0$, $\gamma_1\sim\Gamma(\mu,\mu)$.

\emph{Proof}: By taking the limit $\kappa\rightarrow 0$ in eq.~(\ref{pdf}) and applying the following property
\begin{equation}
\label{limite1}
\lim_{c\rightarrow 0}~_p\mathcal{F}_q\left(a_1\ldots a_p; b_1\ldots b_q; c z\right)=1,
\end{equation}
we obtain the pdf in eq.~(\ref{pdf_Gamma}). Notice that eq.~(\ref{limite1}) can be carried out by simply exploiting the series expression of the hypergeometric function of scalar argument, where the first term has the unit value and the rest of the terms are powers of the scalar argument \cite[eq. (13.1.2)]{Abramowitz}, so that they become zero when taking the limit.

We give the following interpretation about Corollary 2. By tending $\kappa\rightarrow 0$, we eliminate all the dominant components of the model, regardless the value of the shadowing parameter $m$, so that we only have scattering components in each cluster, i.e., we obtain a model which follows a Nakagami-$m$ distribution or one of its particular cases, Rayleigh or one-sided Gaussian, depending on the value of $\mu$. 

\subsection{$\eta$-$\mu$ distribution and particular cases}

The Nakagami-$q$ (Hoyt) and the $\eta$-$\mu$ distributions are employed in non-homogeneous propagation conditions environments, where the scattering model is {non-uniform} and can be modeled by elliptical (or non-circularly symmetric) random variables. At first glance, such scenario does not seem to fit with the $\kappa$-$\mu$ shadowed fading model. However, we can give a different interpretation to the cluster components of the physical model in eq.~(\ref{model_propuesto}): they can be interpreted as a set of uniform scattering waves with random {averages}. These random fluctuations in the average, which are different for each cluster, are responsible for modeling the non-homogeneity of {the environment} and ultimately lead to breaking the circular symmetry of {the scattering model}. {We must note that a similar connection was inferred in \cite{Romero-Javi}, where the \mbox{Nakagami-$q$} distribution was shown to behave as a Rayleigh distribution with randomly varying average power. We show next how the circular symmetry of the model can be broken  by using the result of Lemma 2.}

\emph{Corollary 3}: Let $\gamma_2=\bar{\gamma}_2\Omega_2/\bar{\Omega}_2$, with $\bar{\gamma}_2=\mathbb{E}[\gamma_2]$, be the instantaneous SNR of the model in eq.~(\ref{model_propuesto}), i.e, ${\gamma_2\sim\mathcal{S}_{\kappa\mu m}(\bar{\gamma}; \kappa, \mu, m)}$. If $m=\mu/2$, $\gamma_2\sim\mathcal{S}_{\eta\mu}(\bar{\gamma}; \frac{1}{2\kappa+1}, \mu/2)$.

\emph{Proof}: When $m=\mu/2$, we can apply in eq.~(\ref{pdf}) the following property \cite[eq. (9.6.47)]{Abramowitz}
\begin{equation}
\label{limite2}
_1\mathcal{F}_1\left(a ; 2a; z\right)=2^{2a-1}\Gamma\left(a+\frac{1}{2}\right)z^{\frac{1}{2}-a}\text{e}^{z/2}I_{a-\frac{1}{2}}\left(\frac{z}{2}\right),
\end{equation}
and so we have the eq.~(\ref{pdf_eta_mu}) after some simple algebraic manipulations.

Hence, we have shown that the $\eta$-$\mu$ fading distribution arises as a particular case of the more general $\kappa$-$\mu$ shadowed model. This is one of the main results in this paper. Notice that when {$m=\mu/2=0.5$} we obtain the Nakagami-$q$ model with shape parameter $q=\sqrt{\frac{1}{2\kappa+1}}$ since $\eta=q^2$ for $\eta$-$\mu$ format 1 \cite{Yacoub}.\footnote{The $\eta$-$\mu$ fading model (format 1) is symmetrical for $\eta\in[0,1]$ and $\eta\in[1,\infty]$. We have $q=\sqrt{\eta}$ or $q=1/\sqrt{\eta}$ depending on the interval.} 

We can give the following interpretation of the result in Corollary 3. { Let us consider that each shadowing component can be expressed as $\widetilde{\xi}_i=x_i\cdot e^{j\phi}$, where $x_i\sim\mathcal{N}(0,1)$, so that $\vert\widetilde{\xi}_i\vert^2\sim\Gamma(1/2,1/2)$. When we consider the simplest case on which the phase of $\widetilde{\xi}_i$ is deterministic and set to zero, then $\widetilde{\xi}_i$ becomes a real Gaussian random variable. Thus, we are adding a real Gaussian random variable to a complex Gaussian random variable, so that the circular symmetry of the model is broken. In the general case of an arbitrary phase for $\widetilde{\xi}_i$, the circular symmetry would be broken in a direction of the complex plane different to the real axis.} {Therefore}, while the $\kappa$-$\mu$ model in Section IV.A is obtained by totally eliminating the randomness of the shadowing component, this is not the case for the $\eta$-$\mu$ fading model.

In turn, the Nakagami-$m$ model can be also deduced from the $\kappa$-$\mu$ shadowed physical model of eq.~(\ref{model_propuesto}) with a similar method, i.e., without eliminating directly the dominant component.

\emph{Corollary 4}: Let $\gamma_2=\bar{\gamma}_2\Omega_2/\bar{\Omega}_2$, with $\bar{\gamma}_2=\mathbb{E}[\gamma_2]$, be the instantaneous SNR of the model in eq.~(\ref{model_propuesto}), i.e, ${\gamma_2\sim\mathcal{S}_{\kappa\mu m}(\bar{\gamma}; \kappa, \mu, m)}$. If $m=\mu$, $\gamma_2\sim\Gamma(\mu, \mu)$.

\emph{Proof}: The result is straightforward by applying
\begin{equation}
\label{limite4}
_1\mathcal{F}_1\Big(a; a; z\Big)=e^z
\end{equation}
and making some algebraic manipulations.

Notice that by setting $m=\mu$, we transform the i.i.d. random dominant components of the eq.~(\ref{model_propuesto}) into scattering components. In fact, since {$m=\mu\cdot\hat{m}$, then we set $\hat{m}=1$} and the $i$-th random dominant component becomes a Gaussian random variable. Thus we are adding two Gaussian random variables together in each cluster, which leads to an equivalent Gaussian random variable, so that the one-sided Gaussian, Rayleigh or Nakagami-$m$ models are obtained depending on the number of clusters $\mu$ considered.

The table I summarizes all the models that are derived from the $\kappa$-$\mu$ shadowed fading model, where the $\kappa$-$\mu$ shadowed model parameters are underlined for the sake of clarity. When the $\kappa$-$\mu$ shadowed parameters are fixed to some specific real positive values or tend to some specific limits, we can obtain all the classic central models, i.e., the Rayleigh, one-sided Gaussian, Nakagami-$q$ and Nakagami-$m$, the classic noncentral Rician fading, and their general counterparts, the Rician shadowed, $\kappa$-$\mu$ and $\eta$-$\mu$ fading models. 

It is remarkable that there are two ways for deriving the one-sided Gaussian, Rayleigh and Nakagami-$m$ models, depending on whether the approaches in Section IV.A or Section IV.B are used.

\begin{table}[!t]
\renewcommand{\arraystretch}{1.7}
\caption{Classic and generalized models derived from the $\kappa$-$\mu$ shadowed fading}
\label{table_1}
\centering
\begin{tabular}
{c|c}
\hline
\hline
Channels  & $\kappa$-$\mu$ Shadowed Parameters\\
\hline
\hline
One-sided Gaussian &  a) \b{$\mu$}~$=0.5$, \b{$\kappa$}~$\rightarrow 0$ \\ & b) \b{$\mu$}~$ =0.5$, \b{$m$}~$=0.5$ \\
\hline
\multirow{2}{*}{Rayleigh} &  a) \b{$\mu$}~$=1$, \b{$\kappa$}~$\rightarrow 0$ \\ & b) \b{$\mu$}~$ =1$, \b{$m$}~$=1$ \\
\hline
\multirow{2}{*}{Nakagami-$m$} &  a) \b{$\mu$}~$=m$, \b{$\kappa$}~$\rightarrow 0$ \\ & b) \b{$\mu$}~$ =m$, \b{$m$}~$=m$\\
\hline
Nakagami-$q$ (Hoyt)  & \b{$\mu$}~$=1$, \b{$\kappa$}~$=(1-q^2)/2q^2$, \b{$m$}~=~0.5\\  
\hline
Rician with parameter $K$ & \b{$\mu$}~$ = 1$, \b{$\kappa$}~$=K$, \b{$m$}~$\rightarrow\infty$\\
\hline
$\kappa$-$\mu$ & \b{$\mu$}~$=\mu$, \b{$\kappa$}~$=\kappa$, \b{$m$}~$\rightarrow\infty$\\
\hline
$\eta$-$\mu$ & \b{$\mu$}~$=2\mu$, \b{$\kappa$}~$=(1-\eta)/2\eta$, \b{$m$} $=\mu$\\
\hline
Rician shadowed & \b{$\mu$}~$=1$, \b{$\kappa$}~$=K$, \b{$m$}~$=m$\\
\hline
\hline
\end{tabular}
\end{table}

\section{Application: Ergodic Capacity in the High-SNR Regime}
The characterization of the ergodic channel capacity in fading channels, defined as 
\begin{equation}
\bar{C}[\text{bps/Hz}]\triangleq\int_0^{+\infty}\log_2(1+\gamma)f_\gamma(\gamma)d\gamma,
\end{equation}
where $\gamma$ is the instantaneous SNR at the receiver side, has been a matter of interest for many years \cite{Lee,Gunther,Alouini-Andrea,Sagias}. While for the case of Rayleigh fading it is possible to obtain relatively simple closed-form expressions for the capacity, the consideration of more general fading models \cite{Sagias,daCosta,Celia} leads to very complicated expressions that usually require the use of Meijer G-functions.

In order to overcome the limitation of the exact characterization of $\kappa$-$\mu$ shadowed channel capacity due to its complicated closed-form \cite{Celia}, it seems more convenient to analyze the high-SNR regime. In this situation, the ergodic capacity can be approximated by \cite[eq.~(8)]{Yilmaz}
\begin{equation}
\bar{C}(\bar{\gamma})\vert_{\bar{\gamma}\Uparrow}=\log_2(\bar{\gamma})-L,
\end{equation}
which is asymptotically exact and where $L$ is a constant value independent of the average SNR that can be given by
\begin{equation}
L=-\log_2(e)\frac{d}{dn}AF_{\gamma}^{(n)}\Big\vert_{n=0}.
\end{equation}
In fact, the parameter $L$ can be interpreted as the capacity loss with respect to the additive white Gaussian noise (AWGN) case, since the presence of fading causes $L>0$. When there is no fading, $L=0$ and this reduces to the well-known Shannon result. Using this approach, we derive a simple closed-form expression for the asymptotic capacity of the $\kappa$-$\mu$ shadowed model, which is a new result in the literature.

\emph{Lemma 3}: In the high-SNR regime, the ergodic capacity of a $\kappa$-$\mu$ shadowed channel can be accurately lower-bounded by
\begin{equation}
\label{C_asin}
\bar{C}_{\kappa\mu m}(\bar{\gamma})\vert_{\bar{\gamma}\Uparrow}=\log_2(\bar{\gamma})-L_{\kappa\mu m},
\end{equation}
where $\log_2(\cdot)$ is the binary logarithm, $e$ is the base of the natural logarithm, $\bar{\gamma}$ is the average SNR at the receiver side, i.e. $\bar{\gamma}=\mathbb{E}[\gamma]$, and $L_{\kappa\mu m}$ can be expressed as 
\begin{equation}
\label{t}
\begin{split}
L_{\kappa\mu m}=&-\log_2(e)\psi(\mu)-\log_2\Big(\frac{\mu\kappa+m}{\mu m(1+\kappa)}\Big)\\
&+\log_2(e)\frac{\kappa(\mu-m)}{\mu\kappa+m}\\
&\times_3\mathcal{F}_2\Big(1,1,\mu-m+1;2,\mu+1;\frac{\mu\kappa}{\mu\kappa+m}\Big),
\end{split}
\end{equation}
where $\psi(\cdot)$ is the digamma function \cite[eq.~(6.3.1)]{Abramowitz} and $_3\mathcal{F}_2(\cdot)$ is a generalized hypergeometric function of one scalar argument.

\emph{Proof}: See Appendix A.

Notice that when $\mu=1$, we obtain the ergodic capacity of the Rician shadowed in the high-SNR regime.

As opposed to the exact analysis in \cite{Celia}, which requires for the evaluation of a bivariate Meijer G-function, Lemma 3 provides a very simple closed-form expression for the capacity in the high-SNR regime. More interestingly, since the $\kappa$-$\mu$ and $\eta$-$\mu$ fading channel models are but particular cases of the \mbox{$\kappa$-$\mu$} shadowed distribution, we also obtain the capacity in these scenarios without the need of evaluating Meijer G-functions as in \cite{daCosta}. This is formally stated in the following corollaries. 

\emph{Corollary 5}: In the high-SNR regime, the ergodic capacity of a $\kappa$-$\mu$ channel can be accurately lower-bounded by
\begin{equation}
\bar{C}_{\kappa\mu}(\bar{\gamma})\vert_{\bar{\gamma}\Uparrow}=\log_2(\bar{\gamma})-L_{\kappa\mu},
\end{equation}
where $L_{\kappa\mu}$ can be expressed as
\begin{equation}
\label{t_kmu}
\begin{split}
L_{\kappa\mu}=&-\log_2(e)\psi(\mu)+\log_2(\mu)+\log_2(1+\kappa)\\
&-\kappa\log_2(e)_2\mathcal{F}_2\Big(1,1;2,\mu+1;-\mu\kappa\Big).
\end{split}
\end{equation}

\emph{Proof}: The eq.~(\ref{t_kmu}) is derived by applying the limit $m\rightarrow\infty$ in eq.~(\ref{t}), so that the $_3\mathcal{F}_2(\cdot)$ collapses in a $_2\mathcal{F}_2(\cdot)$ hypergeometric function since
\begin{equation}
\lim_{c\rightarrow\infty}\ _3\mathcal{F}_2(a_1,a_2,c;b_1,b_2;\frac{z}{c})=\ _2\mathcal{F}_2(a_1,a_2;b_1,b_2;z).
\end{equation}

\emph{Corollary 6}: In the high-SNR regime, the ergodic capacity of an $\eta$-$\mu$ channel can be accurately lower-bounded by
\begin{equation}
\bar{C}_{\eta\mu}(\bar{\gamma})\vert_{\bar{\gamma}\Uparrow}=\log_2(\bar{\gamma})-L_{\eta\mu},
\end{equation}
where $L_{\eta\mu}$ can be expressed as
\begin{equation}
\begin{split}
L_{\eta\text{-}\mu}=&-\log_2(e)\psi(2\mu)+\log_2(\mu)+\log_2(1+\eta)\\
&+\log_2(e)\frac{(1-\eta)}{2}\ _3\mathcal{F}_2\big(1,1,\mu+1;2,2\mu+1;1-\eta\big).
\end{split}
\end{equation}

\emph{Proof}: We obtain the $\eta$-$\mu$ asympototic capacity loss from eq.~(\ref{t}) by setting \b{$m$}$=\mu$, \b{$\mu$}$=2\mu$ and \b{$\kappa$}$=\frac{1-\eta}{2\eta}$ as Table I indicates.

Hence, the expressions of the $\kappa$-$\mu$ and $\eta$-$\mu$ asymptotic capacities have been jointly deduced from the result of \mbox{Lemma 3}, which are also new results. Moreover, deriving the asymptotic capacity of the $\kappa$-$\mu$ shadowed has not been harder than deriving the $\kappa$-$\mu$ or $\eta$-$\mu$ asymptotic capacities directly, since the $\kappa$-$\mu$ and $\eta$-$\mu$ moments are expressed, like in the $\kappa$-$\mu$ shadowed case, in terms of a Gauss hypergeometric function \cite{Yacoub}. Thus, we are hitting two (actually three) birds with one stone.

Using the equivalences in Table I, we can obtain even simpler expressions for classic fading models which reduce to existing results in the literature, for Nakagami-m \cite{Yilmaz}, Rician \cite{Rao} and Hoyt \cite{Romero-Javi, RomeroConf}. For the sake of clarity, we omit the straightforward derivations of the rest of asymptotic capacities. Instead, we summarize in Table II their capacity losses with respect to the AWGN channel in the high-SNR regime, where $\Gamma(a,b)$ is the incomplete gamma function \cite[eq. (6.5.3)]{Abramowitz} and $\gamma_e$ is the Euler-Mascheroni constant, i.e., $\gamma_e\approx0.5772$.

\begin{table}[!t]
\renewcommand{\arraystretch}{1.7}
\caption{Ergodic capacity loss in the high-SNR regime for different channels}
\label{table_2}
\centering
\begin{tabular}
{c|c}
\hline
\hline
Channels & Ergodic capacity loss ($L$) [bps/Hz]\\
\hline
\hline
One-sided Gaussian &  $1+\gamma_e\cdot\log_2(e)\approx 1.83$  \\
\hline
Rayleigh &  $\gamma_e\cdot\log_2(e)\approx 0.83$ \\ 
\hline
Nakagami-$m$ &  $\log_2(m)-\log_2(e)\psi(m)$ \\ 
\hline
Nakagami-$q$ (Hoyt) & $1+\gamma_e\cdot\log_2(e)+\log_2\Big(\frac{1+q^2}{(1+q)^2}\Big)$\\
\hline
Rician with parameter $K$ & $\log_2(1+1/K)-\log_2(e)\Gamma(0,K)$\\
\hline
$\kappa$-$\mu$ & $-\log_2(e)\psi(\mu)+\log_2(\mu)+\log_2(1+\kappa)$\\
& $-\kappa\log_2(e)_2\mathcal{F}_2\Big(1,1;2,\mu+1;-\mu\kappa\Big)$\\
\hline
& $-\log_2(e)\psi(2\mu)+\log_2(\mu)$\\
$\eta$-$\mu$ & $+\log_2(1+\eta)+\log_2(e)\frac{(1-\eta)}{2}$\\
& $\times_3\mathcal{F}_2\big(1,1,\mu+1;2,2\mu+1;1-\eta\big)$\\
\hline
& $\gamma_e\cdot\log_2(e)$\\
Rician shadowed & $-\log_2\Big(\frac{K+m}{m(1+K)}\Big)+\log_2(e)\frac{K(1-m)}{K+m}$\\
&$\times_3\mathcal{F}_2\Big(1,1,2-m;2,2;\frac{K}{K+m}\Big)$\\
\hline
\hline
\end{tabular}
\end{table}

\section{Numerical Results}
We now study the evolution of the capacity loss for the \mbox{$\kappa$-$\mu$} shadowed, $\kappa$-$\mu$ and $\eta$-$\mu$ fading models with respect to the AWGN case. In Fig.~\ref{fig:C_clasicos} and Fig.~\ref{fig:C_general}, we plot the ergodic capacity of the classic and generalized fading models, respectively. 

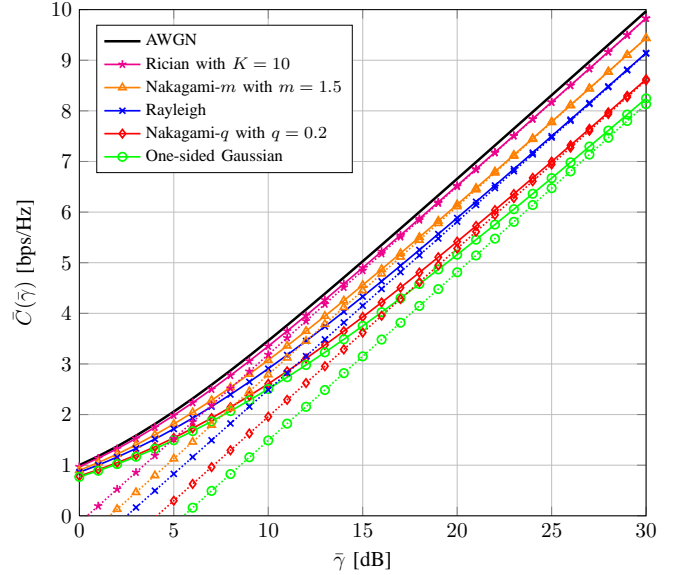
\begin{figure}[!t]
\centering
\pgfplotsset{every axis/.append style={
xtick={0,5,...,30},
xmin={0},
xmax={30},
ytick={0,1,...,10},
ymin={0},
ymax={10},
}}
\begin{tikzpicture}[scale=0.8]
\begin{axis}[
grid=both,
width=11cm,
height=10cm, 
xlabel=$\bar{\gamma}\text{ [dB]}$,
ylabel=$\bar{C}(\bar{\gamma})\text{ [bps/Hz]}$,
legend cell align=left,
legend entries={\text{AWGN},\text{Rician with $K=10$},\text{Nakagami-$m$ with $m=1.5$},\text{Rayleigh}, \text{Nakagami-$q$ with $q=0.2$},\text{One-sided Gaussian}},
legend style={font=\footnotesize, legend pos=north west},
legend columns=1
]

\addplot[color=black, very thick] table[x=x,y=y] {Capacidad_AWGN.dat};

\addplot[color=magenta, thick, mark=star, mark options={solid}] table[x=x,y=y] {Capacidad_Rician_K10.dat};

\addplot[color=orange, thick, mark=triangle, mark options={solid}] table[x=x,y=y] {Capacidad_Nakagami_m1p5.dat};

\addplot[color=blue, thick, mark=x, mark options={solid}] table[x=x,y=y] {Capacidad_Rayleigh.dat};

\addplot[color=red, thick, mark=diamond, mark options={solid}] table[x=x,y=y] {Capacidad_Hoytq0p2.dat};

\addplot[color=green, thick, mark=o, mark options={solid}] table[x=x,y=y] {Capacidad_one_q1e-3.dat};

\addplot[color=blue, thick, densely dotted, mark=x, mark options={solid}] table[x=x,y=y] {Capacidad_asin_kappa0m1e4mu1.dat};

\addplot[color=red, thick, densely dotted, mark=diamond, mark options={solid}] table[x=x,y=y] {Capacidad_asin_kappa12m0p5mu1_eqHoytq0p2.dat};

\addplot[color=magenta, thick, densely dotted, mark=star, mark options={solid}] table[x=x,y=y] {Capacidad_asin_kappa10m1e4mu1_eqRicianK10.dat};

\addplot[color=green, thick, densely dotted, mark=o, mark options={solid}] table[x=x,y=y] {Capacidad_asin_kappa0m1e4mu0p5_eqOne.dat};

\addplot[color=orange, thick, densely dotted, mark=triangle, mark options={solid}] table[x=x,y=y] {Capacidad_asin_kappa0m1e4mu1p5_eqNakagami_m10.dat};

\end{axis}
\end{tikzpicture}
\caption{Comparison of classic channel ergodic capacities with their asymptotic values in the high-SNR regime.}
\label{fig:C_clasicos}
\end{figure}
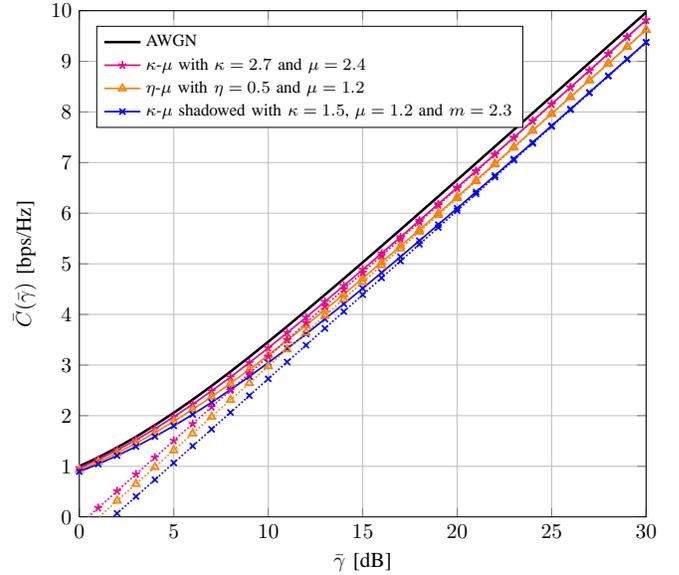
\begin{figure}[!t]
\centering
\pgfplotsset{every axis/.append style={
xtick={0,5,...,30},
xmin={0},
xmax={30},
ytick={0,1,...,10},
ymin={0},
ymax={10},
}}
\begin{tikzpicture}[scale=0.8]
\begin{axis}[
grid=both,
width=11cm,
height=10cm, 
xlabel=$\bar{\gamma}\text{ [dB]}$,
ylabel=$\bar{C}(\bar{\gamma})\text{ [bps/Hz]}$,
legend cell align=left,
legend entries={\text{AWGN},\text{$\kappa$-$\mu$ with $\kappa=2.7$ and $\mu=2.4$},\text{$\eta$-$\mu$ with $\eta=0.5$ and $\mu=1.2$},\text{$\kappa$-$\mu$ shadowed with $\kappa=1.5$, $\mu=1.2$ and $m=2.3$}},
legend style={font=\footnotesize, legend pos=north west},
legend columns=1
]

\addplot[color=black, very thick] table[x=x,y=y] {Capacidad_AWGN.dat};

\addplot[color=magenta, thick, mark=star, mark options={solid}] table[x=x,y=y] {Capacidad_kappa2p7mu2p4.dat};

\addplot[color=orange, thick, mark=triangle] table[x=x,y=y] {C_eta0p5mu2p4.dat};

\addplot[color=blue, thick, mark=x] table[x=x,y=y] {Capacidad_kappa1p5m2p3mu1p2.dat};

\addplot[color=orange, thick, densely dotted, mark=triangle, mark options={solid}] table[x=x,y=y] {Capacidad_asin_eta0p5mu1p2.dat};

\addplot[color=blue, thick, densely dotted, mark=x, mark options={solid}] table[x=x,y=y] {Capacidad_asin_kappa1p5m2p3mu1p2_eqkappamushadowed.dat};

\addplot[color=magenta, thick, densely dotted, mark=star, mark options={solid}] table[x=x,y=y] {Capacidad_asin_kappa2p7m1e4mu2p4eqkappamu.dat};

\end{axis}
\end{tikzpicture}
\caption{Comparison of generalized channel ergodic capacities with their asymptotic values in the high-SNR regime.}
\label{fig:C_general}
\end{figure}

We observe that all the models converge accurately to their asymptotic capacity values, remaining below the Shannon limit, i.e, the capacity of the AWGN channel. Therefore, the asymptotic ergodic capacity expression derived in Lemma 3 for the $\kappa$-$\mu$ shadowed model is here validated for the one-sided Gaussian, Rayleigh, Nakagami-$m$, Nakagami-$q$, Rician, Rician shadowed, $\kappa$-$\mu$ and $\eta$-$\mu$ ergodic capacities in the high-SNR regime.

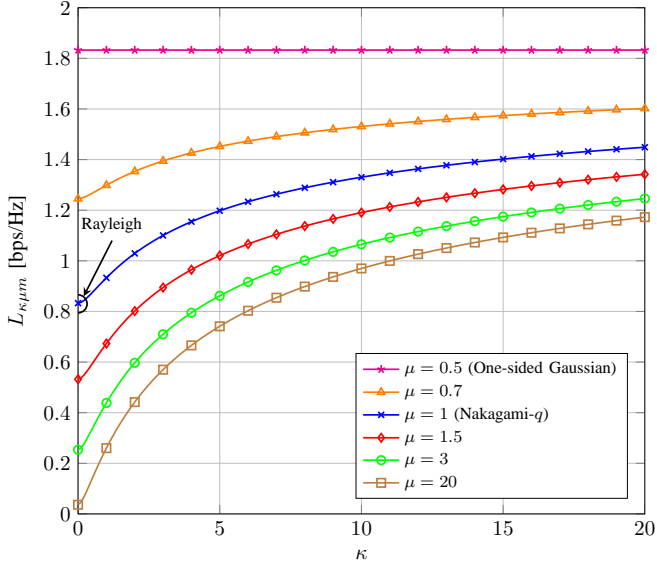
\begin{figure}[!t]
\centering
\pgfplotsset{every axis/.append style={
xtick={0,5,...,20},
xmin={0},
xmax={20},
ymin={0},
ymax={2},
ytick={0,0.2,...,2}
}}
\begin{tikzpicture}[scale=0.8]
\begin{axis}[
grid=both,
width=11cm,
height=10cm, 
xlabel=$\kappa$,
ylabel=$L_{\kappa\mu m}\text{ [bps/Hz]}$,
legend cell align=left,
legend entries={$\mu=0.5$ (One-sided Gaussian),$\mu=0.7$, $\mu=1$ (Nakagami-$q$),$\mu=1.5$,$\mu=3$,$\mu=20$},
legend style={font=\footnotesize, legend pos=south east},
legend columns=1
]

\addplot[color=magenta, thick, mark=star,mark repeat=10] table[x=x,y=y] {Tparam_mu0p5m0p5.dat};

\addplot[color=orange, thick, mark=triangle,mark repeat=10] table[x=x,y=y] {Tparam_mu0p7m0p5.dat};

\addplot[color=blue, thick,mark=x, mark repeat=10] table[x=x,y=y] {Tparam_mu1m0p5.dat};

\addplot[color=red, thick,mark=diamond,mark repeat=10] table[x=x,y=y] {Tparam_mu1p5m0p5.dat};

\addplot[color=green, thick,mark=o,mark repeat=10] table[x=x,y=y] {Tparam_mu3m0p5.dat};

\addplot[color=brown, thick,mark=square,mark repeat=10] table[x=x,y=y] {Tparam_mu20m0p5.dat};

\draw [stealth-,thick] (axis cs:0.2,0.86) -- (axis cs:1.2,1.08) node[above]{\footnotesize Rayleigh};

\draw[thick] (axis cs:0,0.83) circle (0.15cm);

\end{axis}
\end{tikzpicture}
\caption{Evolution of the $\kappa$-$\mu$ shadowed ergodic capacity loss in the high-SNR regime for fixed $m=0.5$.}
\label{fig:m0p5}
\end{figure}

In Figs.~\ref{fig:m0p5}-\ref{fig:m20} we show the evolution of the $\kappa$-$\mu$ shadowed asymptotic capacity loss when $m$ grows. When the shadowing cannot be negligible, i.e, for $m \leq 3$ which corresponds to Figs.~\ref{fig:m0p5}-\ref{fig:m3}, having more power in the dominant components does not always improve the ergodic capacity, but sometimes raises considerably the capacity loss, especially for a great number of clusters. When $m\geq 20$, the shadowing can be neglected and the model actually tends to the $\kappa$-$\mu$ fading, where an increase in the power of the dominant components is obviously favorable for the channel capacity. Therefore, receiving more power through the dominant components does not always increase the capacity in the presence of shadowing. {In fact, we observe two different behaviors in the capacity loss evolution with respect to the parameter $\kappa$. For $m<\mu$, increasing the parameter $\kappa$ is detrimental for the capacity. Conversely, when $m>\mu$ the capacity is improved as $\kappa$ is increased, i.e. in the presence of a stronger LOS component. In the limit case of $m=\mu$, we see that the capacity loss is independent of $\kappa$}. We also see that the capacity loss decreases as $\mu$ grows, since having a larger number of clusters reduces the fading severity of the small-scale propagation effects.

\begin{figure}[!t]
\centering
\pgfplotsset{every axis/.append style={
xtick={0,5,...,20},
xmin={0},
xmax={20},
ymin={0},
ymax={2},
ytick={0,0.2,...,2}
}}
\begin{tikzpicture}[scale=0.8]
\begin{axis}[
grid=both,
width=11cm,
height=10cm, 
xlabel=$\kappa$,
ylabel=$L_{\kappa\mu m}\text{ [bps/Hz]}$,
legend cell align=left,
legend entries={$\mu=0.5$,$\mu=0.7$, $\mu=1$ (Rayleigh),$\mu=1.5$,$\mu=3$,$\mu=20$},
legend style={font=\footnotesize, legend pos=north east},
legend columns=1
]

\addplot[color=magenta, thick, mark=star,mark repeat=10] table[x=x,y=y] {Tparam_mu0p5m1.dat};

\addplot[color=orange, thick, mark=triangle,mark repeat=10] table[x=x,y=y] {Tparam_mu0p7m1.dat};

\addplot[color=blue, thick,mark=x, mark repeat=10] table[x=x,y=y] {Tparam_mu1m1.dat};

\addplot[color=red, thick,mark=diamond,mark repeat=10] table[x=x,y=y] {Tparam_mu1p5m1.dat};

\addplot[color=green, thick,mark=o,mark repeat=10] table[x=x,y=y] {Tparam_mu3m1.dat};

\addplot[color=brown, thick,mark=square,mark repeat=10] table[x=x,y=y] {Tparam_mu20m1.dat};

\draw [stealth-,thick] (axis cs:0.25,1.83) -- (axis cs:1,1.83) node[right]{\footnotesize one-sided Gaussian};

\draw[thick] (axis cs:0,1.83) circle (0.15cm);
\end{axis}
\end{tikzpicture}
\caption{Evolution of the $\kappa$-$\mu$ shadowed ergodic capacity loss in the high-SNR regime for fixed $m=1$.}
\end{figure}
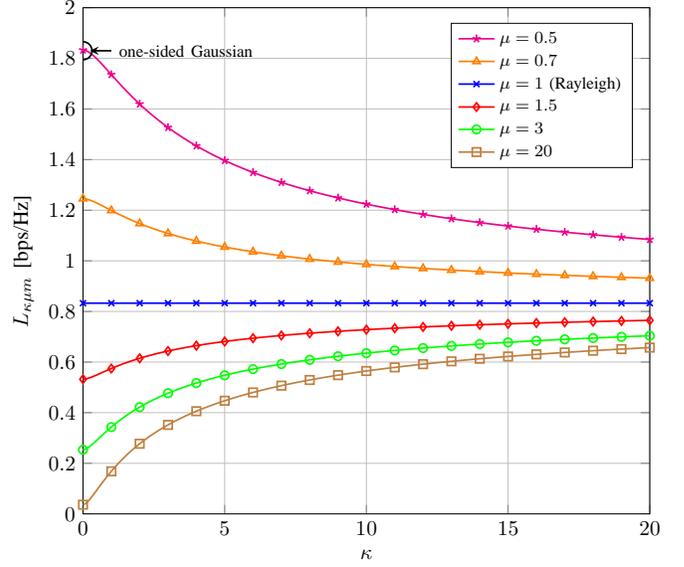
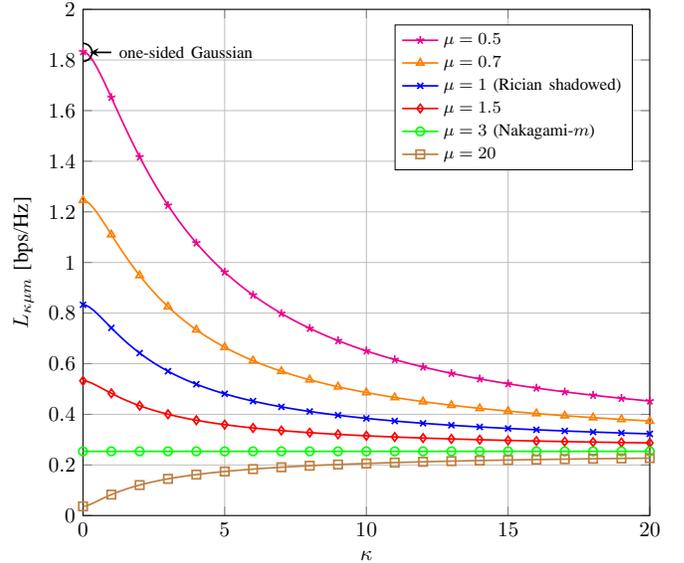
\begin{figure}[!t]
\centering
\pgfplotsset{every axis/.append style={
xtick={0,5,...,20},
xmin={0},
xmax={20},
ymin={0},
ymax={2},
ytick={0,0.2,...,2}
}}
\begin{tikzpicture}[scale=0.8]
\begin{axis}[
grid=both,
width=11cm,
height=10cm, 
xlabel=$\kappa$,
ylabel=$L_{\kappa\mu m}\text{ [bps/Hz]}$,
legend cell align=left,
legend entries={$\mu=0.5$,$\mu=0.7$, $\mu=1$ (Rician shadowed),$\mu=1.5$,$\mu=3$ (Nakagami-$m$),$\mu=20$},
legend style={font=\footnotesize, legend pos=north east},
legend columns=1
]

\addplot[color=magenta, thick, mark=star,mark repeat=10] table[x=x,y=y] {Tparam_mu0p5m3.dat};

\addplot[color=orange, thick, mark=triangle,mark repeat=10] table[x=x,y=y] {Tparam_mu0p7m3.dat};

\addplot[color=blue, thick,mark=x, mark repeat=10] table[x=x,y=y] {Tparam_mu1m3.dat};

\addplot[color=red, thick,mark=diamond,mark repeat=10] table[x=x,y=y] {Tparam_mu1p5m3.dat};

\addplot[color=green, thick,mark=o,mark repeat=10] table[x=x,y=y] {Tparam_mu3m3.dat};

\addplot[color=brown, thick,mark=square,mark repeat=10] table[x=x,y=y] {Tparam_mu20m3.dat};

\draw [stealth-,thick] (axis cs:0.25,1.83) -- (axis cs:1,1.83) node[right]{\footnotesize one-sided Gaussian};

\draw[thick] (axis cs:0,1.83) circle (0.15cm);

\end{axis}
\end{tikzpicture}
\caption{Evolution of the $\kappa$-$\mu$ shadowed ergodic capacity loss in the high-SNR regime for fixed $m=3$.}
\label{fig:m3}
\end{figure}
\begin{figure}[!t]
\centering
\pgfplotsset{every axis/.append style={
xtick={0,5,...,20},
xmin={0},
xmax={20},
ymin={0},
ymax={2},
ytick={0,0.2,...,2}
}}
\begin{tikzpicture}[scale=0.8]
\begin{axis}[
grid=both,
width=11cm,
height=10cm, 
xlabel=$\kappa$,
ylabel=$L_{\kappa\mu m}\text{ [bps/Hz]}$,
legend cell align=left,
legend entries={$\mu=0.5$,$\mu=0.7$, $\mu=1$ (Rician shadowed),$\mu=1.5$,$\mu=3$,$\mu=20$  (Nakagami-$m$)},
legend style={font=\footnotesize, legend pos=north east},
legend columns=1
]

\addplot[color=magenta, thick, mark=star,mark repeat=10] table[x=x,y=y] {Tparam_mu0p5m20.dat};

\addplot[color=orange, thick, mark=triangle,mark repeat=10] table[x=x,y=y] {Tparam_mu0p7m20.dat};

\addplot[color=blue, thick,mark=x, mark repeat=10] table[x=x,y=y] {Tparam_mu1m20.dat};

\addplot[color=red, thick,mark=diamond,mark repeat=10] table[x=x,y=y] {Tparam_mu1p5m20.dat};

\addplot[color=green, thick,mark=o,mark repeat=10] table[x=x,y=y] {Tparam_mu3m20.dat};

\addplot[color=brown, thick,mark=square,mark repeat=10] table[x=x,y=y] {Tparam_mu20m20.dat};

\draw [stealth-,thick] (axis cs:0.25,1.83) -- (axis cs:1,1.83) node[right]{\footnotesize one-sided Gaussian};

\draw[thick] (axis cs:0,1.83) circle (0.15cm);
\end{axis}
\end{tikzpicture}
\caption{Evolution of the $\kappa$-$\mu$ shadowed ergodic capacity loss in the high-SNR regime for fixed $m=20$.}
\label{fig:m20}
\end{figure}
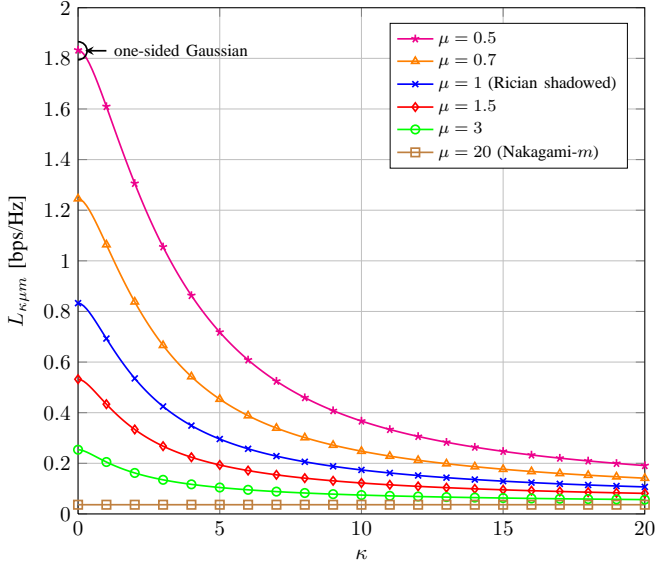

We have also marked in Figs.~\ref{fig:m0p5}-\ref{fig:m20} some models that can be deduced from the $\kappa$-$\mu$ shadowed model. In fact, we can see them in the different legends and also at some specific points rounded by a circle in different curves.

Finally, Fig.~\ref{fig:kappa-mu} and Fig.~\ref{fig:eta-mu} depict the asymptotic ergodic capacity loss for the $\kappa$-$\mu$ and $\eta$-$\mu$ fading models respectively. We observe that Fig.~\ref{fig:kappa-mu} is quite similar to Fig.~\ref{fig:m20} because, as mentioned before, the $\kappa$-$\mu$ shadowed model with $m\geq 20$ can be approximated by the $\kappa$-$\mu$ fading model. In Fig. 8, we see that, regardless the number of clusters $\mu$, there is a minimum in the channel capacity loss at $\eta=1$ which divides in two symmetric parts the fading behavior as expected. It also noticeable that in Fig.~\ref{fig:eta-mu} we have specified the limit cases for $\eta\rightarrow 0$ and $\eta\rightarrow\infty$. In fact, when $\mu=0.5$, the $\eta$-$\mu$ collapses into the one-sided Gaussian model for $\eta=0$ or $\eta\rightarrow\infty$, whereas for $\eta=1$ it collapses into the Rayleigh model. In turn, when $\mu=1$, the $\eta$-$\mu$ is reduced to the Rayleigh case for $\eta=0$ or $\eta\rightarrow\infty$. This is shown in the figure by including also the Rayleigh and one-sided Gaussian capacity loss values with horizontal dotted and dashed lines respectively.  

\section{Conclusions}
We have proved that the $\kappa$-$\mu$ shadowed model unifies the $\kappa$-$\mu$ and $\eta$-$\mu$ fading distributions. { By a novel physical interpretation of the shadowing in the dominant components, we have shown that the $\kappa$-$\mu$ shadowed model can also be employed in non-homogeneous environments, which gives the $\kappa$-$\mu$ shadowed distribution a stronger flexibility to model different propagation conditions than other alternatives, when operating in wireless environments. Thus, the $\kappa$-$\mu$ shadowed model unifies all the classic fading models, i.e., the one-sided Gaussian, Rayleigh, Nakagami-$m$, Nakagami-$q$ and Rician fading channels, and their generalized counterparts, the $\kappa$-$\mu$, \mbox{$\eta$-$\mu$} and Rician shadowed fading models}. Using this connection, simple {new} closed-form expressions have been deduced to evaluate the ergodic capacity in the high-SNR regime for the $\kappa$-$\mu$ shadowed, and hence for the $\kappa$-$\mu$ and \mbox{$\eta$-$\mu$} fading models, giving us clear insights into the contribution of the fading parameters on the capacity improvement or degradation.

As a closing remark, one can think of whether the name of $\kappa$-$\mu$ shadowed distribution is still appropriate for this model, since its flexibility transcends the original characteristics presented in \cite{Paris}.
\begin{figure}[!t]
\centering
\pgfplotsset{every axis/.append style={
xtick={0,5,...,20},
xmin={0},
xmax={20},
ymin={0},
ymax={2},
ytick={0,0.2,...,2}
}}
\begin{tikzpicture}[scale=0.8]
\begin{axis}[
grid=both,
width=11cm,
height=10cm, 
xlabel=$\kappa$,
ylabel=$L_{\kappa\mu}\text{ [bps/Hz]}$,
legend cell align=left,
legend entries={$\mu=0.5$,$\mu=0.7$, $\mu=1$ (Rician),$\mu=1.5$,$\mu=3$,$\mu=20$},
legend style={font=\footnotesize, legend pos=north east},
legend columns=1
]

\addplot[color=magenta, thick, mark=star,mark repeat=10] table[x=x,y=y] {Tparam_mu0p5.dat};

\addplot[color=orange, thick, mark=triangle,mark repeat=10] table[x=x,y=y] {Tparam_mu0p7.dat};

\addplot[color=blue, thick,mark=x, mark repeat=10] table[x=x,y=y] {Tparam_mu1.dat};

\addplot[color=red, thick,mark=diamond,mark repeat=10] table[x=x,y=y] {Tparam_mu1p5.dat};

\addplot[color=green, thick,mark=o,mark repeat=10] table[x=x,y=y] {Tparam_mu3.dat};

\addplot[color=brown, thick,mark=square,mark repeat=10] table[x=x,y=y] {Tparam_mu20.dat};

\draw [stealth-,thick] (axis cs:0.25,1.83) -- (axis cs:1,1.83) node[right]{\footnotesize one-sided Gaussian};

\draw [stealth-,thick] (axis cs:0.1,0.86) -- (axis cs:2.5,1.23) node[right]{\footnotesize Rayleigh};

\draw[thick] (axis cs:0,1.83) circle (0.15cm);

\draw[thick] (axis cs:0,0.83) circle (0.15cm);

\end{axis}
\end{tikzpicture}
\caption{Evolution of the $\kappa$-$\mu$ ergodic capacity loss in the high-SNR regime.}
\label{fig:kappa-mu}
\end{figure}
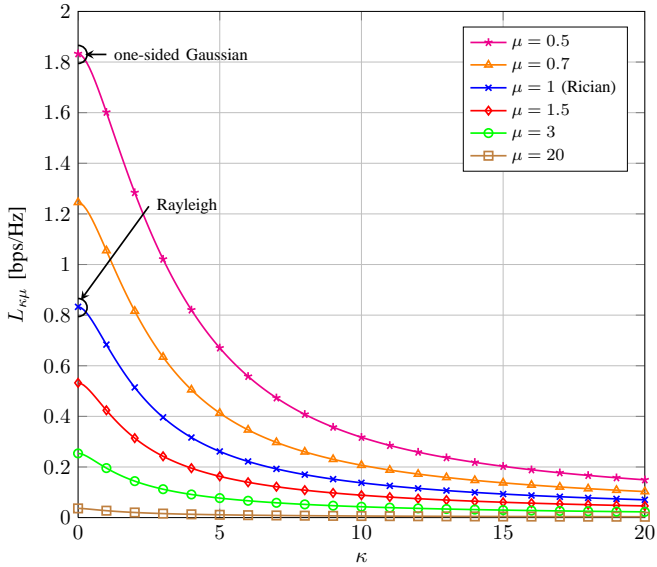
\begin{figure}[!t]
\centering
\pgfplotsset{every axis/.append style={
xmode=log,
xmin={0.001},
xmax={1000},
ymin={0},
ymax={2},
ytick={0,0.2,...,2}
}}
\begin{tikzpicture}[scale=0.8]
\begin{axis}[
grid=both,
width=11cm,
height=10cm, 
xlabel=$\eta$,
ylabel=$L_{\eta\mu}\text{ [bps/Hz]}$,
legend cell align=left,
legend style={font=\footnotesize, at={(0.3,0.82)},anchor=west},
legend columns=1
]

\addlegendimage{magenta,mark=star}
\addlegendentry{$\mu=0.5$ (Nakagami-$q$)}
\addlegendimage{orange,mark=triangle}
\addlegendentry{$\mu=0.7$}
\addlegendimage{blue,mark=x}
\addlegendentry{$\mu=1$}
\addlegendimage{red,mark=diamond}
\addlegendentry{$\mu=1.5$}
\addlegendimage{green,mark=o}
\addlegendentry{$\mu=3$}
\addlegendimage{brown,mark=square}
\addlegendentry{$\mu=20$}

\addplot[color=magenta, thick, mark=star,mark indices={1,3,10,12,19,21,28,48,115,120,127,129,136}] table[x=x,y=y] {Tparam0a1000_etamu0p5.dat};

\addplot[color=orange, thick, mark=triangle,mark indices={1,3,10,12,19,21,28,48,115,120,127,129,136}] table[x=x,y=y] {Tparam0a1000_etamu0p7.dat};

\addplot[color=blue, thick,mark=x, mark indices={1,3,10,12,19,21,28,48,115,120,127,129,136}] table[x=x,y=y] {Tparam0a1000_etamu1.dat};

\addplot[color=red, thick,mark=diamond,mark indices={1,3,10,12,19,21,28,48,115,120,127,129,136}] table[x=x,y=y] {Tparam0a1000_etamu1p5.dat};

\addplot[color=green, thick,mark=o,mark indices={1,3,10,12,19,21,28,48,115,120,127,129,136}] table[x=x,y=y] {Tparam0a1000_etamu3.dat};

\addplot[color=brown, thick,mark=square,mark indices={1,3,10,12,19,21,28,48,115,120,127,129,136}] table[x=x,y=y] {Tparam0a1000_etamu20.dat};

\draw [stealth-,thick] (axis cs:1,0.86) -- (axis cs:1,1.05) node[above]{\footnotesize Rayleigh};

\draw [stealth-,thick] (axis cs:1000,0.86) -- (axis cs:300,0.86);

\draw [stealth-,thick] (axis cs:0.001,0.86) -- (axis cs:0.004,0.86);

\draw [stealth-,thick] (axis cs:1000,1.78) -- (axis cs:300,1.71) ;

\draw [stealth-,thick] (axis cs:0.001,1.78) -- (axis cs:0.004,1.71);

\draw[thick] (axis cs:1,0.83) circle (0.15cm);

\draw[very thick, densely dashed] (axis cs:0.001,1.83) -- (axis cs:1000,1.83);

\draw[very thick, dotted] (axis cs:0.001,0.83) -- (axis cs:1000,0.83);

\end{axis}
\end{tikzpicture}
\caption{Evolution of the $\eta$-$\mu$ ergodic capacity loss in the high-SNR regime. The Rayleigh and the one-sided Gaussian capacity loss particular cases are also included in horizontal dotted and dashed lines respectively.}
\label{fig:eta-mu}
\end{figure}
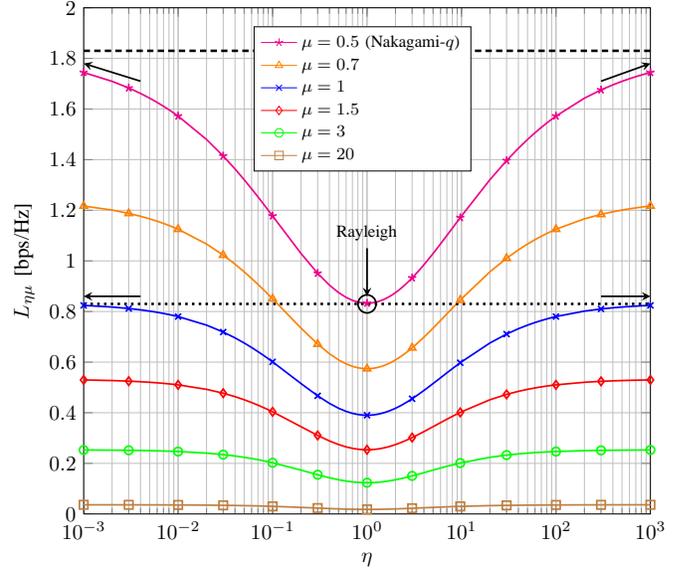
\begin{appendices}
\section{Ergodic capacity of the $\kappa$-$\mu$ shadowed fading in the high-SNR regime}
In the high-SNR regime, it is well-known that the ergodic capacity can be lower-bounded by the eq.~(\ref{C_asin}), where the parameter $L_{\kappa\mu m}$ is related with the $n$th-order derivative of the amount of fading $AF_{\gamma}^{(n)}$, such as  \cite{Yilmaz}
\begin{equation}
L_{\kappa\mu m}=-\log_2(e)\frac{d}{dn}AF_{\gamma}^{(n)}\Big\vert_{n=0}.
\end{equation}
Thus, thanks to eq.~(\ref{AF}), we first have to compute the moments of the SNR at the receiver side before deriving $L_{\kappa\mu m}$, i.e.
\begin{equation}
\begin{split}
\mathbb{E}[\gamma^n]\triangleq&\int_{0}^{+\infty}\gamma^nf_{\gamma}(\gamma)d\gamma\\
=&\frac{\mu^\mu m^m(1+\kappa)^\mu}{\Gamma(\mu)\bar{\gamma}^\mu(\mu\kappa+m)^m}\\&\times\int_{0}^{+\infty}\gamma^{\mu+n-1}\text{e}^{-\frac{\mu(1+\kappa)\gamma}{\bar{\gamma}}}\\
&\times_1\mathcal{F}_1\left(m,\mu;\frac{\mu^2\kappa(1+\kappa)}{\mu\kappa+m}\frac{\gamma}{\bar{\gamma}}\right)d\gamma.
\end{split}
\end{equation}
Observing that the remaining integral correspond to a Laplace transform evaluated in $s=\frac{\mu(1+\kappa)}{\bar{\gamma}}$, we then have \cite[eq.~(4.23.17)]{Erdelyi}
\begin{equation}
\begin{split}
\label{moments}
\mathbb{E}[\gamma^n]=&\frac{\Gamma(\mu+n)}{\Gamma(\mu)}\frac{\bar{\gamma}^n m^m(1+\kappa)^{-n}}{\mu^n(\mu\kappa+m)^m}\\
&\times_2\mathcal{F}_1\left(m;\mu+n;\mu;\frac{\mu\kappa}{\mu\kappa+m}\right),
\end{split}
\end{equation}
where $_2\mathcal{F}_1(\cdot)$ is the Gauss hypergeometric function of scalar argument \cite[eq. (15.1.1)]{Abramowitz}.

By making a well-known transformation involving the arguments of the Gauss hypergeometric function \cite[eq.~(15.3.3)]{Abramowitz}, we obtain
\begin{equation}
\begin{split}
\label{moments2}
\mathbb{E}[\gamma^n]=&\frac{\Gamma(\mu+n)}{\Gamma(\mu)}\Big(\frac{\mu\kappa+m}{\mu m(1+\kappa)}\Big)^n\\
&\times\ _2\mathcal{F}_1\left(\mu-m,-n;\mu;\frac{\mu\kappa}{\mu\kappa+m}\right).
\end{split}
\end{equation}
The amount of fading is then deduced by using the product rule in eq.~(\ref{moments2}) with the Gauss hypergeometric function expressed in series form \cite[eq.~(15.1.1)]{Abramowitz}. As the derivative of a Pochhammer symbol can be given by a difference of digamma functions $\psi(\cdot)$, we obtain
\begin{equation}
\begin{split}
\frac{d}{dn}AF_{\gamma}^{(n)}=&\frac{\Gamma(\mu+n)}{\Gamma(\mu)}\Big(\frac{\mu\kappa+m}{\mu m(1+\kappa)}\Big)^n\\
&\times\Big\{\Big[\psi(\mu+n)+\log\Big(\frac{\mu\kappa+m}{\mu m(1+\kappa)}\Big)\Big]\\
&\times\ _2\mathcal{F}_1\Big(\mu-m,-n;\mu;\frac{\mu\kappa}{\mu\kappa+m}\Big)\\
&-\sum_{r=1}^{+\infty}\frac{(\mu-m)_r(-n)_r}{(\mu)_r}(\psi(-n+r)-\psi(-n))\\
&\times\frac{\big(\frac{\mu\kappa}{\mu\kappa+m}\big)^r}{r!}\Big\},
\end{split}
\end{equation} 
where the infinite sum starts at $r=1$ because the first term equals zero.
Setting the moments order $n=0$, we get
\begin{equation}
\begin{split}
L_{\kappa\mu m}=&-\log_2(e)\frac{d}{dn}AF_{\gamma}^{(n)}\Big\vert_{n=0}\\
=&-\log_2(e)\psi(\mu)-\log_2\Big(\frac{\mu\kappa+m}{\mu m(1+\kappa)}\Big)\\
&+\log_2(e)\sum_{r=1}^{+\infty}\frac{(\mu-m)_r(1)_{r-1}}{(\mu)_r}\frac{\big(\frac{\mu\kappa}{\mu\kappa+m}\big)^r}{r!}.
\end{split}
\end{equation}
By applying some algebraic manipulations, we finally obtain 
\begin{equation}
\begin{split}
L_{\kappa\mu m}=&-\log_2(e)\psi(\mu)-\log_2\Big(\frac{\mu\kappa+m}{\mu m(1+\kappa)}\Big)\\
&+\log_2(e)\frac{\kappa(\mu-m)}{\mu\kappa+m}\\
&\times\sum_{r=1}^{+\infty}\frac{(\mu-m+1)_{r-1}(1)_{r-1}(1)_{r-1}}{(\mu+1)_{r-1}(2)_{r-1}}\frac{\big(\frac{\mu\kappa}{\mu\kappa+m}\big)^{r-1}}{(r-1)!},
\end{split}
\end{equation}
where the infinite sum can be expressed in terms of the generalized hypergeometric function $_3\mathcal{F}_2(\cdot)$ and so we have the result of eq.~(\ref{t}). Notice that this result gives a simple new expression for the derivative of the Gauss hypergeometric funcion $_2\mathcal{F}_1(a,b;c,z)$ with respect to $a$ or $b$, when this same parameter $a$ or $b$ equals zero. In fact, this derivative is expressed in terms of the generalized hypergeometric function $_3\mathcal{F}_2(\cdot)$ instead of in terms of a Kamp\'e de F\'eriet function as proposed in \cite{Ancarani}.
\end{appendices}

\end{document}